\documentclass[12pt]{iopart}
\usepackage{graphicx}
\begin{document}
\title[Cross-correlation between UHECR arrival distribution and LSS]{Cross-Correlation between UHECR Arrival Distribution and Large-Scale Structure}
\author{Hajime Takami$^{1,2}$, Takahiro Nishimichi$^1$, Kazuhiro Yahata$^1$, and Katsuhiko Sato$^{2,3}$}
\address{$^1$ Department of Physics, School of Science, the University of Tokyo, 7-3-1 Hongo, Bunkyo-ku, Tokyo 113-0033, Japan}
\address{$^2$ Institute for the Physics and Mathematics of the Universe, the University of Tokyo, 5-1-5, Kashiwanoha, Kashiwa, Chiba 277-8582, Japan}
\address{$^3$ Department of Physics, School of Science and Engineering, 
Meisei University, 2-1-1 Hodokubo, Hino-shi, Tokyo 191-8506, Japan}
\ead{takami@utap.phys.s.u-tokyo.ac.jp}

\begin{abstract}
We investigate correlation between the arrival directions 
of ultra-high-energy cosmic rays (UHECRs) 
and the large-scale structure (LSS) of the Universe 
by using statistical quantities 
which can find the angular scale of the correlation. 
The Infrared Astronomical Satellite 
Point Source Redshift Survey (IRAS PSCz) catalog of galaxies 
is adopted for LSS. 
We find a positive correlation of the highest energy events 
detected by the Pierre Auger Observatory (PAO) 
with the IRAS galaxies inside $z=0.018$ 
within the angular scale of $\sim 15^{\circ}$. 
This positive correlation observed in the southern sky implies 
that a significant fraction of the highest energy events 
comes from nearby extragalactic objects. 
We also analyze the data of the Akeno Giant Air Shower Array (AGASA) 
which observed the northern hemisphere, 
but the obvious signals of positive correlation with the galaxy distribution 
are not found. 
Since the exposure of the AGASA is smaller than the PAO, 
the cross-correlation in the northern sky should be tested 
using a larger number of events detected in the future. 
We also discuss the correlation using the all-sky combined data sets 
of both the PAO and AGASA, 
and find a significant correlation within $\sim 8^{\circ}$. 
These angular scales can constrain several models of 
intergalactic magnetic field. 
These cross-correlation signals can be well reproduced by a source model 
in which the distribution of UHECR sources is related to the IRAS galaxies. 
\end{abstract}
\maketitle

\section{Introduction} \label{introduction}

The origin of ultra-high-energy cosmic rays (UHECRs) above $10^{19}$ eV 
has been an open problem in astroparticle physics. 
The highly isotropic distribution of their arrival directions 
and the deflection angles of UHECRs estimated in the Galaxy 
have indicated the extragalactic origin of them. 
In 1999, Akeno Giant Air Shower Array (AGASA) found 
a small-scale anisotropy of the arrival distribution of UHECRs 
above $4 \times 10^{19}$ eV within $2.5^{\circ}$, 
which is comparable with its angular resolution \cite{takeda99}. 
The small-scale anisotropy can be interpreted as 
a signal of point-like sources.

Motivated by the AGASA result, many researchers have tested 
correlation between the arrival directions of UHECRs 
and the positions of several classes of astrophysical objects. 
The correlation with BL Lac objects was discussed by 
using the AGASA and Yakutsk data \cite{tinyakov01,tinyakov02,gorbunov02}. 
The authors of Refs. \cite{smialkowski02,singh04} studied 
the correlation with infrared galaxies 
using {\it Infrared Astronomical Satellite Point Source Redshift Survey} 
(IRAS PSCz) catalog \cite{saunders00}, 
and discussed the possibility that 
the AGASA events below $\sim 10^{20}$eV come from luminous infrared galaxies. 
In Ref. \cite{hague07}, 
the correlation with nearby active galactic nuclei (AGNs) 
listed in the {\it Rossi X-ray Timing Explorer} (RXTE) catalog of AGNs 
was investigated and 
no significant correlation was found. 
The authors of Ref. \cite{gorbunov05} made a comprehensive study 
of correlation with various classes of powerful extragalactic sources 
and found that BL Lac objects 
and unidentified gamma-ray sources might be correlated 
with the AGASA and Yakutsk data. 
The spatial correlation with the supergalactic plane 
has also been discussed and 
positive signals have been obtained 
\cite{stanev95,uchihori00,stanev08}. 
On the other hand, 
High Resolution Fly's Eye (HiRes) reported no significant signal 
for any point-like sources \cite{abbasi05,abbasi07} and 
the correlation with BL Lac objects studied by Refs. 
\cite{tinyakov01,tinyakov02,gorbunov02} was not supported 
by the HiRes data \cite{abbasi06}. 
Theoretically, 
several authors predicted the anisotropy of UHECR arrival distribution 
in future observations 
when UHECR source distribution follows the large-scale structure (LSS) 
of the Universe \cite{waxman97,yoshiguchi03b,takami08,kashti08}. 
Cosmic rays above $8 \times 10^{19}$ eV lose their energies rapidly 
by interactions with cosmic microwave background (CMB) 
and therefore they cannot reach the earth from distant sources, 
typically above 100 Mpc (Greisen-Zatsepin-Kuz'min (GZK) mechanism 
\cite{greisen66,zatsepin66}). 
In other words, 
the GZK mechanism predicts positional correlation 
between such highest energy cosmic rays and nearby UHECR sources 
if the Universe is not magnetized. 
We predicted such positional correlation within a few degree scale 
even taking into account a structured intergalactic magnetic field (IGMF), 
which reproduces the observed structure of local Universe, 
and 5 yrs observation of the Pierre Auger Observatory (PAO) 
will unveil the local UHECR source distribution \cite{takami08}.

Recently, the PAO has reported correlation between the arrival directions of 
its 27 events above $5.7 \times 10^{19}$eV and the positions of nearby active 
galactic nuclei (AGNs) with 3$\sigma$ confidence level \cite{pao07a,pao08}. 
This report showed that the isotropic distribution of the arrival directions 
of the highest energy cosmic rays is disfavored. 
This was also a clear evidence which indicated directly that UHECR sources are 
extragalactic objects and can be regarded as the start of UHECR astronomy.

However, the interpretation of the PAO results is problematic. 
All of the PAO-correlated AGNs except Centaurus A 
are classified into Seyfert galaxies and 
Low-Ionization Nuclear Emission Regions (LINERs), 
which have much weaker activities than radio-loud AGNs, 
one of plausible candidates of UHECR sources \cite{moskalenko08}. 
Thus, it is an open question whether the PAO-correlated AGNs are 
really UHECR sources.

The PAO adopted the 12th edition of Veron-Cetty \& Veron Catalog (VC catalog) 
of AGNs \cite{veron06} for a correlation analysis. 
This catalog is a compilation of many astronomical catalogs of AGNs 
available in literature 
and therefore AGNs in the VC catalog are selected based on different criteria. 
In order to investigate the nature of UHECR sources, 
the authors of Ref. \cite{george08} analyzed the spatial correlation 
of the PAO events with hard X-ray selected AGNs compiled 
by the Swift satellite, an AGN catalog complied based on a criterion. 
They found a significant correlation by using the 2 dimensional generalization 
of the Kolmogorov-Smilnov test (2D KS test).

More generally, the arrival directions of UHECRs are expected to correlate 
with a galaxy distribution, which is a representative of LSS 
of the Universe, if UHECR sources are some astrophysical objects. 
When a correlation with LSS is investigated, 
a catalog as homogeneous as possible is required. 
Ref. \cite{kashti08} studied correlation between the PAO events 
and LSS using the IRAS PSCz catalog. 
They showed that the PAO data is inconsistent with random distribution 
of UHECR sources at 98\% confidence level 
and favors a modeled source distribution following LSS. 
They analyzed the similarity between the observed arrival distribution 
of UHECRs and mock one from their cosmic ray intensity map constructed 
based on the IRAS catalog using $6^{\circ} \times 6^{\circ}$ angular bins. 
The bin size was determined by considering possible deflections of UHECRs 
by the IGMF in order to avoid the effect of the IGMF. 
Ref. \cite{ghisellini08} discussed the correlation with HI-selected galaxies 
by using 2D KS test and a significant correlation was found. 
The authors interpreted the result as an evidence that spiral galaxies 
are the hosts of UHECR sources and could suggest that 
newly born magnetars are the best candidates of UHECR sources. 
Note that 2D KS test cannot estimate the angular scale 
of a spatial correlation. 
These works are important steps toward the understanding of UHECR sources. 
However, if the angular scale is naturally found by another analysis 
without fixing the scale, 
we can obtain a better evidence of the correlation and 
information on intervening magnetic fields and the composition of UHECRs.

In this study, we investigate spatial correlation between 
the arrival directions of the observed highest energy cosmic rays and 
a galaxy distribution as a representative of LSS 
by using statistical quantities which need not to set 
an artificial angular scale and can estimate the angular scale 
of the correlation. 
We adopt the PAO data for cosmic rays observed in the southern hemisphere 
and the AGASA data for cosmic rays detected in the northern hemisphere. 
The HiRes is a UHECR observatory with the largest exposure 
in the northern hemisphere and 
recently, the HiRes reported no significant correlation with AGNs listed 
in the VC catalog in their data by using the same method 
as the PAO \cite{abbasi08}. 
However, HiRes has not published the detailed data and, 
moreover, the analytical estimation of the apertures of fluorescence detectors 
is more complicated than that of ground-based detectors. 
Thus, we do not consider the HiRes in this paper. 
We also check whether a source model associated with LSS 
can reproduce the cross-correlation signal 
and try to constrain the IGMF strength 
taking into account UHECR propagation in magnetized intergalactic space. 
This approach allows us to analyze the correlation 
taking account of the radial information of the positions of UHECR sources.

This paper is laid out as follows. 
In Section \ref{methods}, 
we explain a galaxy catalog which represents LSS and UHECR samples 
used in this study, and our methods of the statistical analysis. 
In Section \ref{correlation}, 
we calculate cross-correlation function 
introduced in Section \ref{methods} 
between the arrival directions of the observed highest energy cosmic rays 
and the positions of nearby galaxy distribution, 
and discuss correlation with its angular scale. 
In Section \ref{modelp}, 
we construct a simple model of UHECR sources, 
check the reproducibility of the observed cross-correlation 
and try to constrain the IGMF strength. 
Finally, in Section \ref{conclusion}, 
we discuss our results and make conclusions. 

\section{Sample data and Statistical Methods} \label{methods}

\subsection{Galaxy and UHECR Catalogs}

We adopt the IRAS PSCz catalog of galaxies \cite{saunders00}. 
In order to represent the LSS of the Universe by galaxies, 
a uniform catalog of galaxies is appropriate. 
The IRAS catalog is a flux-limited catalog to contain 14,677 galaxies 
with redshift and covers a large fraction of the entire sky ($\sim 84\%$). 
Almost all of the uncovered area is around the Galactic plane. 
Thus, this is a good catalog to study correlation between 
the arrival distribution of UHECRs and LSS.

As the observed data of the arrival directions of UHECRs, 
we adopt 
published data by the southern site of the PAO \cite{pao08}, 
which contains 27 events above $5.7 \times 10^{19}$ eV, 
and data published by the AGASA with 
57 events above $4 \times 10^{19}$ eV \cite{hayashida00}. 
The total exposure of the PAO 
is $9.0 \times 10^3~{\rm km}^2~{\rm sr}$, 
and that of AGASA, 
which is the largest ground array in the northern hemisphere, 
is $1.3 \times 10^3~{\rm km}^2~{\rm sr}$. 

\begin{figure}[t]
\begin{center}
\rotatebox{-90}{\includegraphics[width=0.40\linewidth]{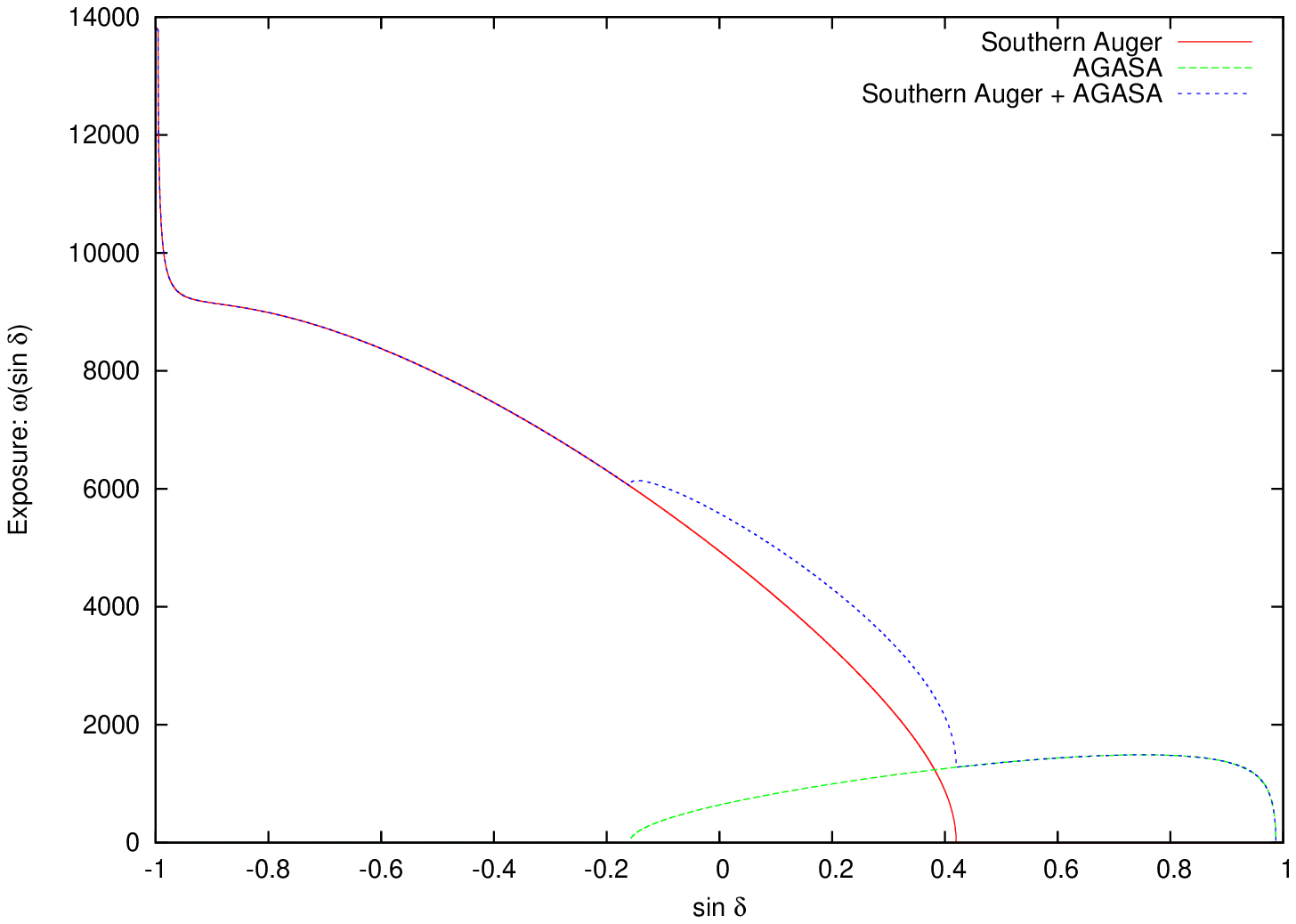}}
\caption{Declination ($\delta$) dependence of the exposures 
of the PAO ({\it red}) and AGASA ({\it green}). 
The total exposures are 9,000 ${\rm km}^2{\rm sr}$ for the PAO 
and $1.3 \times 10^3 {\rm km}^2{\rm sr}$ for the AGASA. 
The combined exposure is also shown ({\it blue}).} 
\label{fig:exposure}
\end{center}
\end{figure}

The apertures of cosmic ray experiments are not uniform, 
but the apertures of ground arrays only depend on 
the declination of the arrival directions of cosmic rays, $\delta$. 
Their exposures, $\omega(\delta)$, 
can be estimated analytically as \cite{sommers01}
\begin{equation}
\omega(\delta) \propto \cos(a_0) \cos(\delta) \sin (\alpha_m) 
+ \alpha_m \sin(a_0) \sin(\delta), 
\end{equation}
where $\alpha_m$ is given by 
\begin{equation}
\alpha_m = \left\{
\begin{array}{ll}
0 & {\rm if}~\xi > 1 \\
\pi & {\rm if}~\xi < -1 \\
\cos^{-1}(\xi) & {\rm otherwise}
\end{array}
\right.
\end{equation}
and 
\begin{equation}
\xi \equiv \frac{\cos(\theta) - \sin(a_0) \sin(\delta)}{\cos(a_0) \cos(\delta)}. 
\end{equation}
Here $a_0$, and $\theta$ are 
the terrestrial latitude of a ground array 
and the zenith angle for an experimental cut. 
These values are $a_0=-35.2^{\circ}$, $\theta=60^{\circ}$ for the PAO 
\cite{pao07a,pao08}, and $a_0=35^{\circ}.47'$, $\theta=45^{\circ}$ 
for the AGASA \cite{takeda99}. 
Fig. \ref{fig:exposure} represents the declination dependence 
of the exposures of both experiments and the combined experiment. 
The total exposure of the AGASA is about 7 times smaller than that of the PAO. 
These dependence of the exposures is taken into account throughout this paper.

\subsection{Estimators of Statistical Quantities} \label{statistics}

A traditional statistical quantity in cosmology 
to investigate similarity between two distribution 
is cross-correlation function \cite{peebles80}. 
This function can be calculated from pair-counts between two distribution 
if the positions of all objects are known on the whole sky. 
Now we would like to calculate the cross-correlation function 
between the arrival directions of UHECRs 
and the positions of the IRAS galaxies. 
As explained above, 
the IRAS PSCz catalog has 16\% of unobserved area ({\it mask}) 
and the apertures of the UHECR observatories depend on 
the declination of the arriving events, 
so that both apertures are not uniform. 
In such case, 
estimators of the cross-correlation function, 
which can infer the cross-correlation function 
in the case of uniform apertures 
have been adopted in observational cosmology.

We adopt 
\begin{equation}
w_{\rm eg}(\theta) = \frac{EG(\theta) - EG'(\theta) - E'G(\theta) + E'G'(\theta)}{E'G'(\theta)}, 
\end{equation}
as the estimator of the cross-correlation function, 
which is a modified version of that used in Ref. \cite{blake06} 
originally suggested for auto-correlation function by Ref. \cite{landy93}. 
$EG(\theta)$ is normalized pair-counts between galaxies in the IRAS catalog 
and cosmic ray events with the separation angle of $\theta$, 
which is obtained by dividing calculated pair-counts by $N_eN_g$, 
where $N_e$ and $N_g$ are the number of events and the number of galaxies, 
respectively. 
$EG'(\theta)$, $E'G(\theta)$, and $E'G'(\theta)$ are the similar pair counts, 
but $G'$ and $E'$ represent galaxies randomly put 
inside the observed sky 
and cosmic ray events 
randomly put with number density proportional to the detector apertures, 
respectively. 
These randomly put galaxies and events 
enable us to correct the effect of non-uniform apertures. 
These galaxies and events must be put with sufficiently large number 
to reflect the apertures. 
We consider 400,000 random distributed galaxies, 
and 200,000 and 400,000 randomly put events 
for the calculation of $w_{\rm eg}(\theta)$ using the PAO or AGASA data 
separately and the combined all-sky data, respectively. 
The width of the angular bin is set to be $1^{\circ}$, 
which is comparable with the angular resolution 
of the PAO \cite{pao07a,pao08}. 
If there is no correlation, $w_{\rm eg}(\theta)$ is equal to zero. 
Positive value of $w_{\rm eg}(\theta)$ corresponds to positive correlation 
and vice versa. 
Thus, we search angular scales in which $w_{\rm eg}(\theta)$ is positive.

We also discuss the auto-correlation of galaxy distribution. 
We adopt an estimator of the auto-correlation function 
suggested by Ref. \cite{landy93}, 
\begin{equation}
w_{\rm gg}(\theta) = \frac{GG(\theta) - 2GG'(\theta) + G'G'(\theta)}{G'G'(\theta)}, 
\end{equation}
where $GG(\theta)$, $GG'(\theta)$, and $G'G'(\theta)$ are 
normalized pair-counts obtained by dividing calculated pair-counts 
by $N_g(N_g-1)/2$, $N_gN'_g$, and $N'_g(N'_g-1)/2$, respectively.

\section{Correlation with Nearby Galaxies} \label{correlation}

In this section, we search cross-correlation signals 
between the PAO and AGASA data, and the positions of the IRAS galaxies. 
The cross-correlation functions of the data defined in Section \ref{statistics} 
are calculated and are compared with that calculated from event distribution 
randomly placed following the apertures of the detectors. 
We adopt the IRAS galaxies within $z=0.018$ as nearby galaxy distribution, 
which is the same criterion as analyses by the PAO \cite{pao07a,pao08}.

\begin{figure}[t]
\begin{center}
\includegraphics[width=0.98\linewidth]{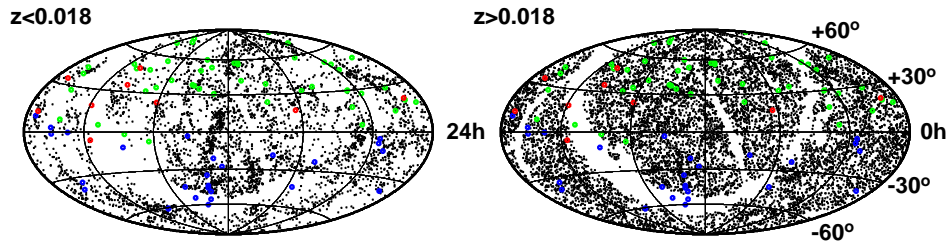}
\caption{Distributions of the IRAS galaxies 
within ({\it left}) and outside ({\it right}) of $z=0.018$ 
with the arrival directions of the highest energy events 
detected by PAO ({\it blue points}) above $5.7 \times 10^{19}$eV 
and AGASA ({\it green and red points}) above $4.0 \times 10^{19}$eV. 
The red points represent the AGASA data above $9.5 \times 10^{19}$eV 
in its energy-reconstruction scale 
and the green points are those below $9.5 \times 10^{19}$eV. }
\label{fig:map}
\end{center}
\end{figure}

Fig.\ref{fig:map} shows the galaxy distributions of the IRAS catalog 
within ({\it left}) and outside ({\it right}) $z=0.018$ 
with the arrival directions of the highest energy events observed 
by the PAO ($E \geq 5.7 \times 10^{19}$ eV; {\it blue}) 
and AGASA ($E \geq 4.0 \times 10^{19}$ eV; {\it green}, 
$E \geq 9.5 \times 10^{19}$ eV; {\it red}) in equatorial coordinates. 
The classification of the AGASA data is for convenience 
when we discuss the combined data of the PAO and AGASA. 
The numbers of galaxies within and outside $z=0.018$ are 4419 and 10258, 
respectively. 
In the left figure, several nearby clusters of galaxies can be found. 
In the PAO data, 
we can find that about 1/3 of the 27 events seem to be close to 
the direction of Centaurus cluster. 
On the other hand, 
we cannot see obvious correlation with galaxies by eye in the AGASA data. 
In the results of both observatories, 
no UHE events have been detected at the highest energies 
in the direction of Virgo cluster, 
whose coordinate is at $\sim (12{\rm h}30{\rm m}, +10^{\circ})$, 
as pointed out by Refs. \cite{gorbunov07,gorbunov08}.  
In the right figure, 
it is difficult to find galaxy clusters obviously 
because of the large number of galaxies projected to 2 dimensional space 
and the small number density of the IRAS galaxies at distant Universe 
due to the selection effect. 
Instead, the IRAS mask can be clearly seen. 
Several events of the PAO and AGASA are in the mask, 
but nevertheless we use all the events for analysis.

\begin{figure}[t]
\begin{center}
\includegraphics[width=0.95\linewidth]{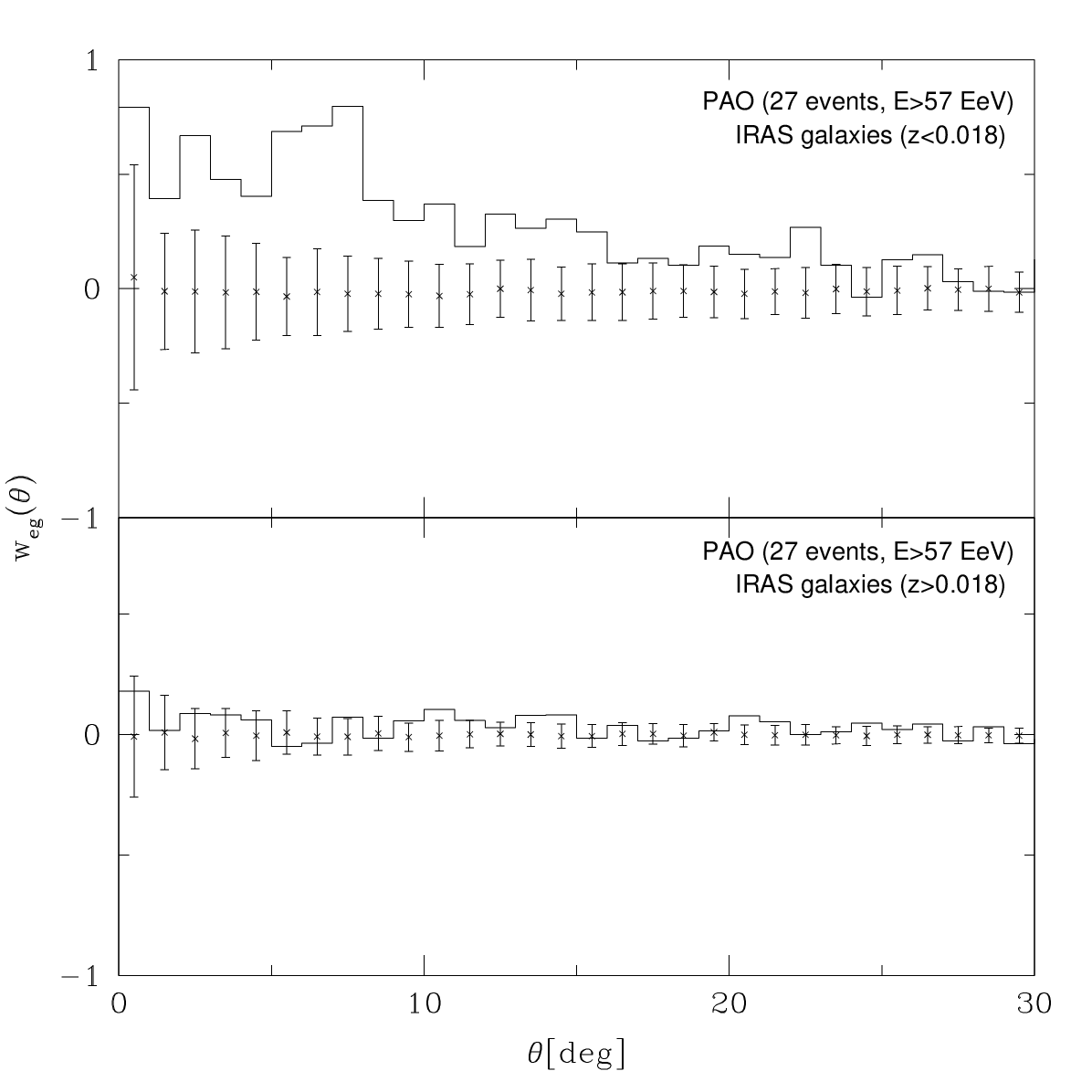}
\caption{Cross-correlation functions, $w_{\rm eg}(\theta)$, 
between the highest energy events detected by the PAO 
and the IRAS galaxies within ({\it upper panel}) 
and outside $z=0.018$ ({\it lower panel}). 
The histograms are $w_{\rm eg}(\theta)$ calculated from the observed data. 
$w_{\rm eg}(\theta)$ calculated from random distribution, 
taking the non-uniform exposure of the PAO into account, 
is also shown with the standard deviation 
due to the finite number of events.}
\label{fig:ccor_pao}
\end{center}
\end{figure}

Fig.\ref{fig:ccor_pao} shows cross-correlation functions 
between the PAO events and the IRAS galaxies 
within ({\it upper panel}) and outside ({\it lower panel}) $z=0.018$. 
For comparison, 
we also show cross-correlation functions calculated 
from randomly distributed cosmic rays with the same number of events, 
taking the anisotropic exposure of the PAO into account. 
The random event generations are performed 100 times, and 
we plot the mean values by dots and standard deviations by error bars. 
In each angular bin, 
the value of $w_{\rm eg}(\theta)$ for random events 
in each realization approximately follows 
the Gaussian distribution. 
Thus, the standard deviations can be regarded 
as 1 $\sigma$ statistical errors. 
We can check that the effects of the non-uniform exposures 
of the PAO and the IRAS mask are neatly corrected 
from the fact that all averaged values of $w_{\rm eg}(\theta)$ 
calculated from randomly distributed mock events are consistent with zero.

In the case of $z \leq 0.018$, 
the cross-correlation function of the PAO data 
is well beyond the error bars of random event distribution within $15^{\circ}$. 
Thus, we can find that the PAO data has significantly positive correlation 
with local galaxy distribution within $z=0.018$ 
at the angular scale of $\leq 15^{\circ}$. 
This means that UHECR deflection angles with respect to the line of sight 
to the sources, $\theta_{\rm obs}$, are less than $15^{\circ}$. 
On the other hand, in the lower panel, 
the PAO data is consistent with random distribution. 
Since the auto-correlation of the outside galaxies has a sharp peak 
at a small angular scale as explained afterward in this section, 
the consistency shows that there is no correlation 
between the PAO data and the outside galaxy distribution. 
These facts represent that a significant fraction of the 27 PAO events 
are injected from astrophysical sources within $z=0.018$.

\begin{figure}[t]
\begin{center}
\includegraphics[width=0.48\linewidth]{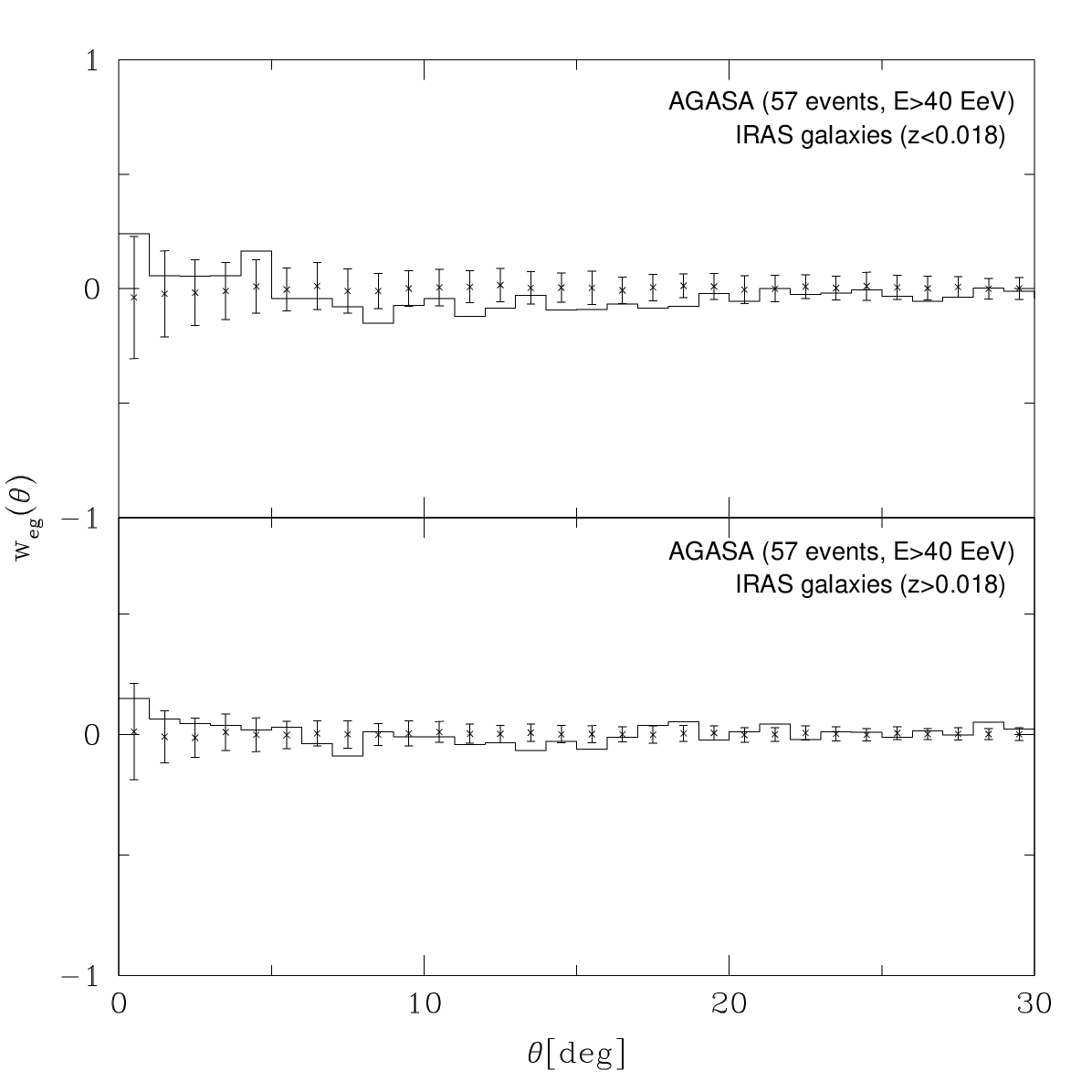} \hfill
\includegraphics[width=0.48\linewidth]{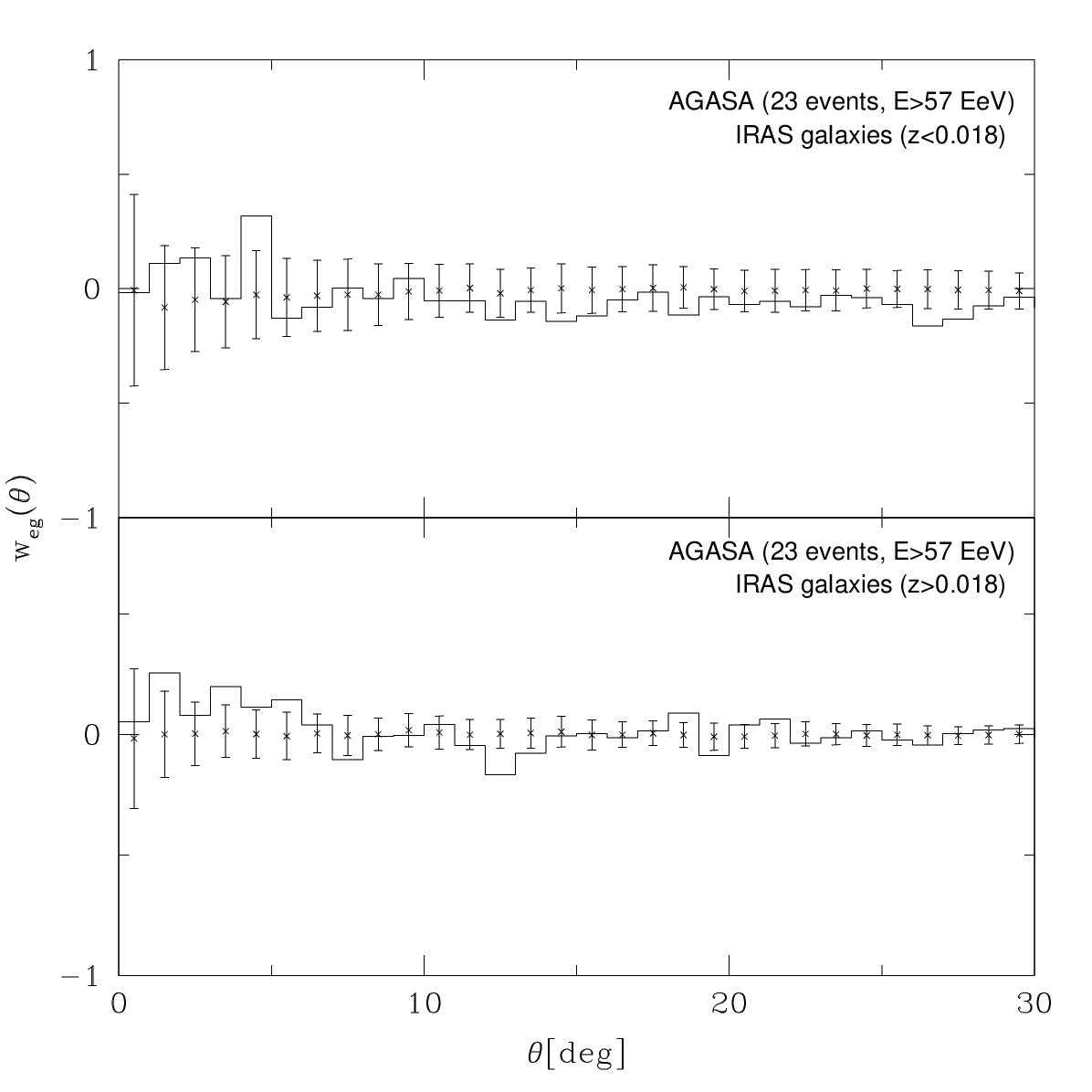}
\caption{The same figures as Fig.\ref{fig:ccor_pao}, but for the AGASA data. 
The threshold energies of events used are 
$4.0 \times 10^{19}$ ({\it left}) and $5.7 \times 10^{19}$ eV ({\it right}) 
respectively.}
\label{fig:ccor_agasa}
\end{center}
\end{figure}

Fig.\ref{fig:ccor_agasa} is the same figure as Fig.\ref{fig:ccor_pao}, 
but for the AGASA data. 
The left panel shows the cross-correlation functions 
calculated from all published data of the AGASA ($E > 40$ EeV, 57 events) 
and the right panel is those calculated using the AGASA data only above 57 EeV 
in the energy scale of the AGASA (23 events). 
In both panel, we cannot find the obvious signals of positive correlation 
like Fig.\ref{fig:ccor_pao}.

\begin{figure}[t]
\begin{center}
\rotatebox{-90}{\includegraphics[width=0.34\linewidth]{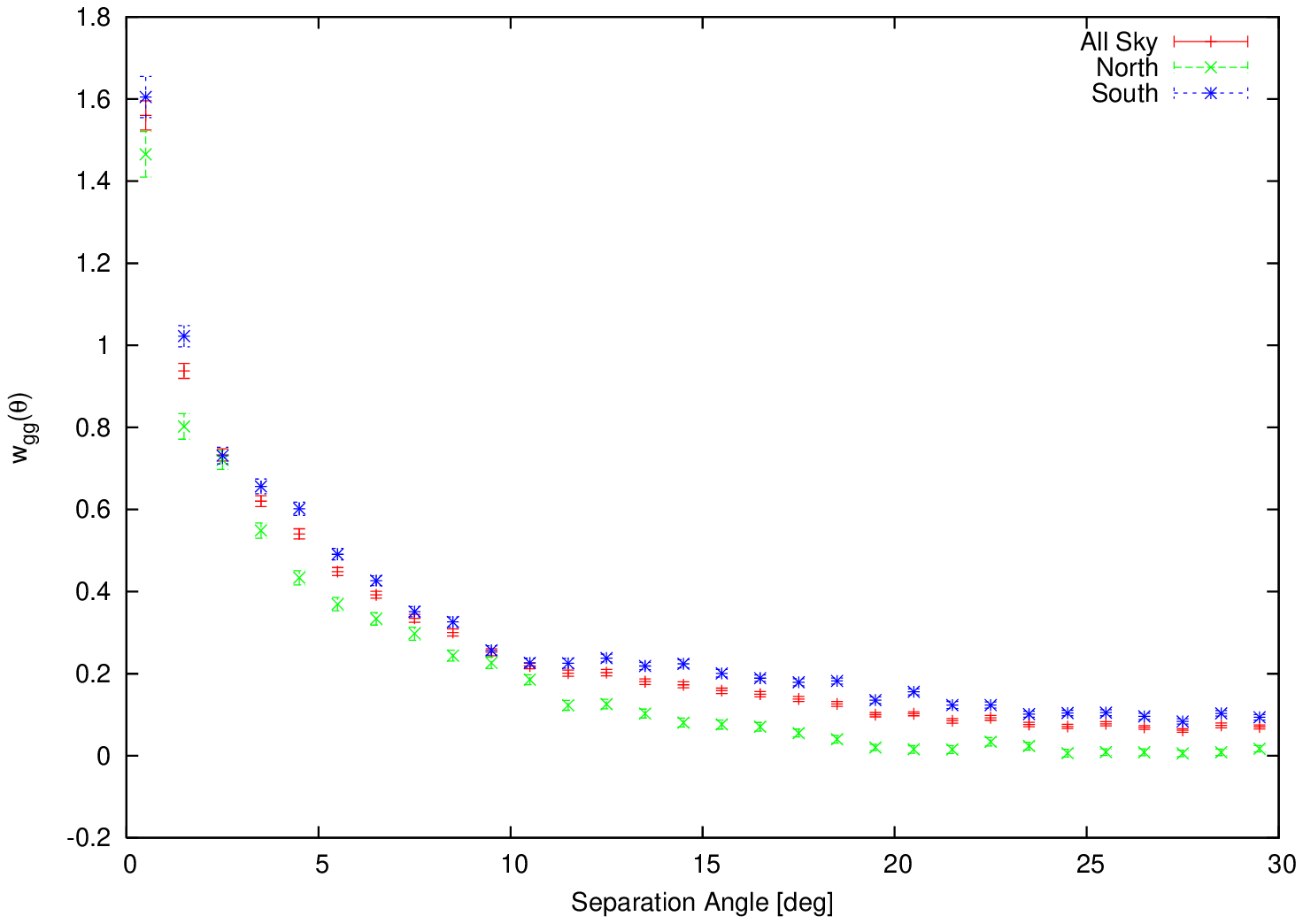}} \hfill
\rotatebox{-90}{\includegraphics[width=0.34\linewidth]{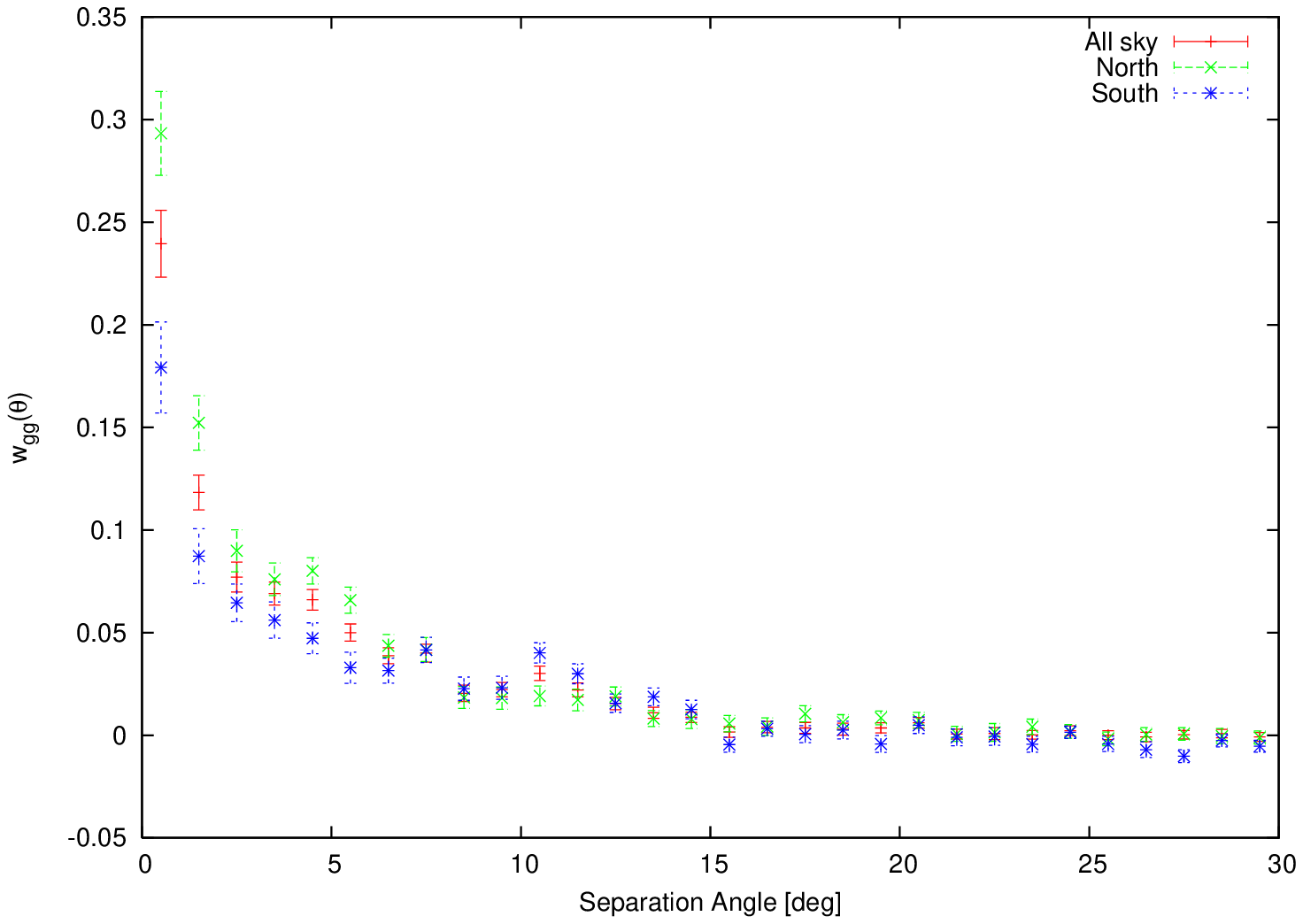}}
\caption{Auto-correlation functions of the IRAS galaxies within $z=0.018$ 
({\it left}) and beyond $z=0.018$ ({\it right}) 
in the northern sky ({\it green}), the southern sky ({\it blue}), 
and all sky ({\it red}).}
\label{fig:auto}
\end{center}
\end{figure}

We also consider the auto-correlation of galaxies 
and compare the angular scale of the cross-correlation 
to that of the auto-correlation. 
Fig. \ref{fig:auto} shows the auto-correlation functions of galaxies 
within ({\it left}) and beyond ({\it right}) $z=0.018$ in the northern 
sky ({\it green}), southern sky ({\it blue}), and all sky ({\it red}). 
The error bars are estimated by 100 realizations of the isotropic 
distribution of galaxies with the same numbers. 
Both panels show the enhancement of the auto-correlations 
at small angular scales. 
Since the projection of galaxies $z > 0.018$ is not isotropic, 
the cross-correlation test with galaxies beyond $z=0.018$ is justified.

The left panel shows that the angular scale of auto-correlations 
of the nearby galaxies is $\sim 10^{\circ}$. 
This scale is comparable with the angular scale of 
the cross-correlation in the southern sky. 
This fact shows that UHECR sources in the southern hemisphere 
are compatible with being distributed with the distribution of galaxies. 
However, we should notice that we cannot distinguish 
following two possibilities: 
UHECR sources themselves are distributed over the galaxy distribution, 
or the IGMF in the surrounding of UHECR sources deflects 
the trajectories of UHECRs and diffuse their arrival directions 
by the angular scale. 
In any case, we can conclude that 
the deflection angles of UHECRs are less than $\sim 10^{\circ}$.

In the northern sky, cross-correlation between UHECRs 
and nearby galaxies was not found as seen in Fig.\ref{fig:ccor_agasa} 
while galaxies cluster at the small angular scale. 
We could suggest 2 possibilities for the lack of positive correlation. 
One is due to the energy-scale of the AGASA. 
The number of events with energies above 57 EeV (23 events) is comparable with 
that of the PAO (27 events) while the exposure of the AGASA is 7 times smaller 
than that of the PAO. 
This might be due to the difference of the energy-scale 
between the observations. 
If we require that the AGASA and PAO should observe the same UHECR spectrum, 
the energy-scale of the AGASA is expected to be shifted to lower energies 
and vice versa for the PAO. 
Thus, the 23 events include events with energies below 57 EeV. 
Since UHECRs with lower energies can arrive at the Earth 
from more distant sources, cross-correlation with nearby galaxies is weakened. 
Although galaxies outside $z = 0.018$ cluster at a small angular scale, 
the signal is smaller than in the case of $z \leq 0.018$ 
and thus the obvious signals of cross-correlation are not found 
for $z > 0.018$. 
The deflections by the IGMF, which is neglected in this section, 
are also expected to affect the lack of cross-correlation for $z > 0.018$ 
because of the relatively long propagation lengths of UHECRs. 
However, it is difficult to check this possibility 
because the difference of the energy-scale originates 
from not statistical errors but systematic errors.

The other is the lack of UHECR sources in nearby Universe in the northern sky, 
that is due to the difference between the northern and southern UHE Universe. 
However, it may be unlikely that there is the lack of UHECR sources 
despite the enhancement of the auto-correlation of nearby galaxies 
at a small angular scale if UHECR sources are astrophysical objects. 
This possibility can be discussed with a numerical simulation. 
In the next section, 
we discuss the reproducibility of the cross-correlation, 
especially for the AGASA, 
if UHECR sources are distributed to being comparable with 
the local galaxy distribution.

\section{Model Prediction} \label{modelp}

If UHECR sources are astrophysical objects, 
the arrival distribution of UHECRs is expected to be related to 
the local distribution of the sources \cite{yoshiguchi03b,takami08} 
or generally to galaxies in local universe \cite{waxman97,kashti08}. 
In this section, 
we test the reproducibility of cross-correlation function 
calculated from the observed data 
by simulating the arrival distribution of UHECRs 
based on a simple source model related to the galaxy distribution. 
The simulation takes the propagation of UHECRs 
in uniformly magnetized intergalactic space into account. 
The composition of UHECRs is assumed to be purely protons 
according to an implication based on the small deflections of UHECRs 
in the Galactic magnetic field (GMF) by the PAO \cite{pao08}. 
Note that there are also claims that the composition 
of the highest energy cosmic rays includes a significant fraction 
of heavier components 
though there is uncertainty on hadronic interaction models 
in extensive air shower \cite{unger07,engel07,glushkov07}.

The mock UHECR source distribution is constructed by a method 
used in Ref.\cite{takami06} based on the IRAS PSCz catalog of galaxies. 
This method allows us 
to construct source distribution related to the galaxy distribution. 
In the IRAS catalog, 
the distances of some nearby galaxies 
estimated by Ref.\cite{rowan_robinson88} are given. 
The distances of the other galaxies are determined 
from their recession velocities listed in the IRAS catalog, 
assuming the $\Lambda$CDM cosmology with 
$H_0 = 71$ km s$^{-1}$ Mpc$^{-1}$, 
$\Omega_m = 0.3$, and $\Omega_{\Lambda} = 0.7$, 
which are the Hubble constant at present, 
matter density and cosmological constant normalized 
by the critical density, respectively. 
The structured source distribution is adopted within 200 Mpc 
because such sources can contribute to about 90\% of 
UHECR flux above $6 \times 10^{19}$ eV. 
For the outside of 200 Mpc, source distribution is assumed to be isotropic. 
We adopt $10^{-4}~{\rm Mpc}^{-3}$ as the number density of UHECR sources, 
which can well reproduce the small-scale anisotropy 
observed by the PAO \cite{takami08c,cuoco08}. 
The UHECR ejection power is assumed to be identical over all sources. 
We consider 100 mock source distributions with the same source number density 
changing the random seed. 
Each source is selected with the same weight, 
not dependent on any property of galaxies like luminosity.

Methods for calculating the propagation of UHE protons and 
their arrival distribution follow those of Ref.\cite{yoshiguchi03}. 
UHE protons lose their energies by interactions 
with the CMB \cite{berezinsky88,yoshida93} 
and their trajectories are deflected by the IGMF during their propagation. 
The IGMF is assumed to be a turbulent field with the Kolmogorov spectrum. 
The coherent length, $l_c$, is also assumed to be 1 Mpc, 
which is much larger than values indicated by Ref. \cite{dolag05} 
and used in Ref. \cite{kashti08}. 
The strength of the IGMF, $B$, is a free parameter since it is poorly known. 
The propagating protons interact with the CMB 
and lose their energies through photopion production 
and Bethe-Heitler pair creation. 
For pair creation, 
we adopt an analytical fitting function given by Ref. \cite{chodorowski92} 
to calculate the energy-loss rate in isotropic photons. 
Photopion production is treated as a stochastic process, 
and the interaction length calculated by an event generator 
SOPHIA \cite{mucke00} is adopted. 
The propagating protons also lose their energies 
through adiabatic energy-loss due to the cosmic expansion, 
but it is neglected in this study 
because this is not important for protons with energies above $10^{19}$ eV. 
We consider UHE protons with $10^{19}$-$10^{22}$ eV. 
5,000 protons are injected in each of 30 energy bins, 
that is 10 bins per decade of energy, 
from a source, and their trajectories and energies are calculated.  
The energies and deflection angles of protons are recorded 
at every 1 Mpc for $d \leq 100$ Mpc while at every 10 Mpc for $d > 100$ Mpc 
where $d$ is the distance from the source, 
and their frequency distributions are constructed. 
The frequency distributions at a certain distance 
can be regarded as those of arriving protons from a source with that distance. 
Based on these frequency distributions, a given source distribution, 
and a given injection spectrum, 
the arrival distribution of mock UHE protons can be simulated. 
A power-law spectrum for UHECR injection, $E^{-\alpha}$, is assumed. 
We set $\alpha = 2.6$ which is well reproduce the observed UHECR spectra 
above $10^{19}$ eV. 
Note that all results are almost independent of the detailed value of $\alpha$, 
as long as $\alpha$ is around 2-3, 
because a spectral shape at the highest energy level 
is dominantly determined by the GZK mechanism. 
We also consider the errors in arrival direction of the UHECR experiments. 
The angular distribution of the error is assumed 
to be a 2 dimensional Gaussian distribution with zero mean and 
the standard deviation equal to the detector resolution, 
which is $1.0^{\circ}$ for the PAO and $1.8^{\circ}$ for the AGASA.

\begin{figure}[t]
\begin{center}
\includegraphics[width=0.95\linewidth]{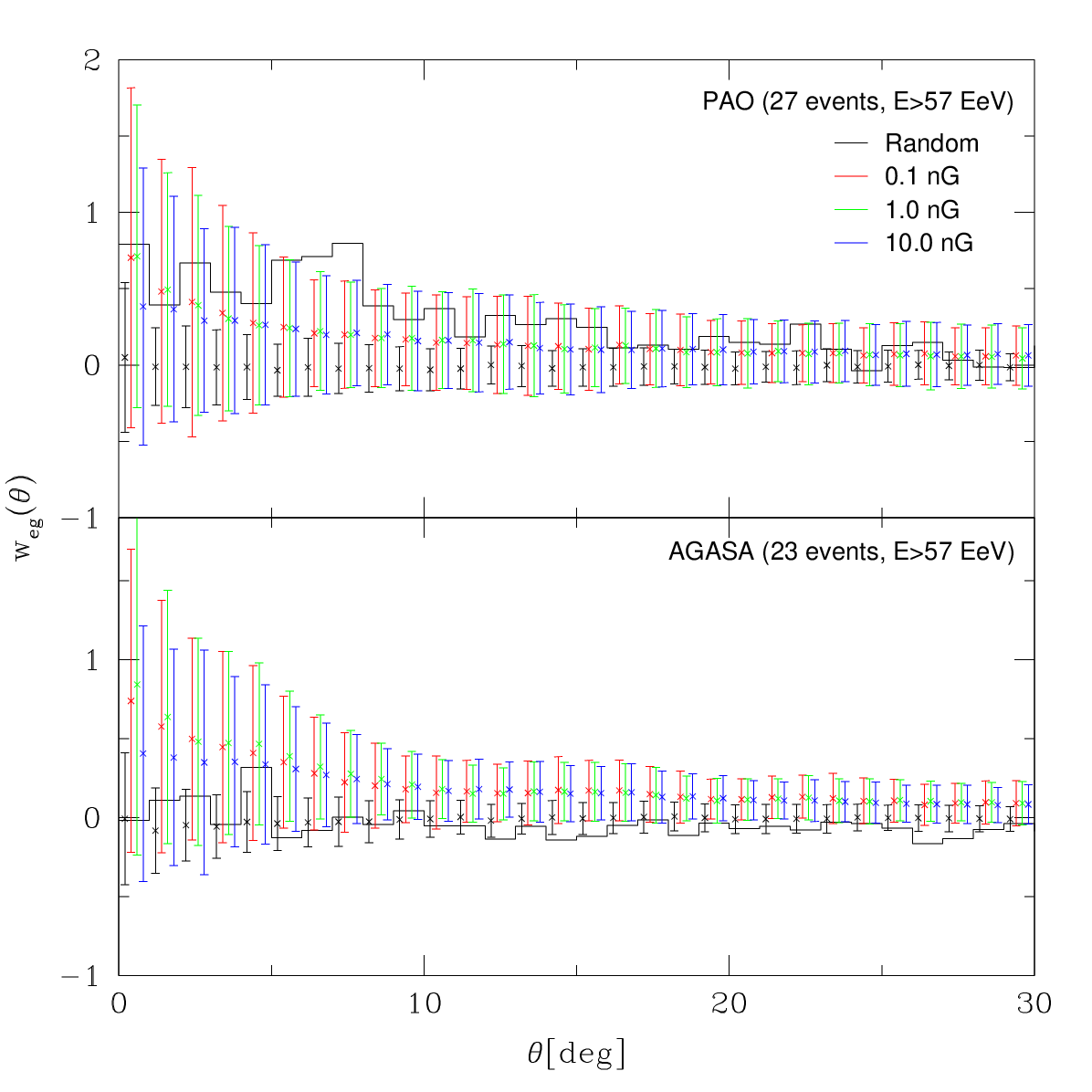}
\caption{Cross-correlation functions predicted by our source model. 
The color points and error bars are 
in the case of $B=$0.1 ({\it red}), 1.0 ({\it green}), 
and 10.0 nG ({\it blue}) respectively. 
The histograms, and black points and error bars are the same 
as the upper panels of Fig.\ref{fig:ccor_pao} and Fig.\ref{fig:ccor_agasa}.}
\label{fig:model1}
\end{center}
\end{figure}

Fig. \ref{fig:model1} shows the cross-correlation functions 
of the IRAS galaxies with $z \leq 0.018$ 
and UHE protons simulated for 
$B=0.1$ ({\it red}), $1.0$ ({\it green}), and $10.0$ nG ({\it blue}). 
The error bars represent standard deviations 
which are estimated from event realizations. 
The event realization is performed 1 time for every source distribution. 
Thus, the error bars include the error 
due to not only the finite number of events 
but also the sampling of galaxies in the different realizations. 
The apertures of the PAO ({\it upper panel}) 
and the AGASA ({\it lower panel}) are taken into account. 
The number of events is set to the same 
as the observed data, shown in the figure. 
The histograms and the black error bars are 
the same as in the upper panels of Figs.\ref{fig:ccor_pao} 
and \ref{fig:ccor_agasa}.

For the PAO, the mock data for the 3 different IGMF strengths reproduce 
cross-correlation consistent with the observed data 
within the standard deviations. 
Thus, the distribution of UHECR sources is evenly compatible with 
the local galaxy distribution in the southern hemisphere.

In principle, the cross-correlation of the observed data allows 
constraining the strength of IGMF 
because stronger IGMF deflects more the trajectories of protons 
and loses cross-correlation at a small angular scale. 
As we expect, 
the averages of the cross-correlation functions for $B=10.0$ nG are smaller 
than those for $B=1.0$ nG at a small angular scale in this figure. 
However, unfortunately, the difference between the averages in the 2 models 
is much smaller than the error bars. 
Thus, the discussion on the difference is only qualitative 
due to large error bars and we cannot distinguish the IGMF strength 
with sufficient significance.

It has a reason that the prediction for $B=0.1$nG is almost the same as that 
for $B=1.0$nG though the difference between the two is within the error bars. 
The reason is the finite resolution of cosmic ray arrival directions 
by UHECR observatories, typically $\sim 1^{\circ}$. 
Typical deflection angle of UHE proton is represented as 
\begin{equation}
\theta_{\rm cr} (E,d) \simeq 0.4^{\circ} Z 
\left( \frac{E}{60~{\rm EeV}} \right)^{-1} 
\left( \frac{d}{100~{\rm Mpc}} \right)^{1/2} 
\left( \frac{l_c}{1~{\rm Mpc}} \right)^{1/2} 
\left( \frac{B}{0.1~{\rm nG}} \right), 
\label{deflection}
\end{equation}
where $d$ is the distance of a UHECR source. 
Note that the formula shows a typical angle 
between initial and final velocity of a propagating proton, 
$\theta_{\rm cr}(E,d)$. 
Typically, $\theta_{\rm obs}$ is about a half of 
$\theta_{\rm cr}(E,d)$. 
$\theta_{\rm obs}$ for $B = 0.1 $ nG is smaller than the angular resolution. 
Thus, the IGMF strength less than a few times 0.1 nG cannot be resolved 
by currently operating UHECR observatories. 
Note that it is possible that IGMF with the strength of $\sim 1$ nG 
is not distinguishable if $l_c$ smaller than 1 Mpc is realized in the Universe. 

For the AGASA, the mock data almost reproduce the observational data 
within the standard deviations. 
The cross-correlation function of the AGASA can be reproduced 
by UHECR sources distributed with being comparable to galaxy distribution, 
especially within the angular scale of $\sim 10^{\circ}$. 
Thus, the lack of cross-correlation at a small angular scale 
in the AGASA data is still within errors. 
This means that the lack of UHECR sources in nearby northern sky 
is not always needed at present. 
When we watch cross-correlation functions at a larger angular scale, 
we find that the mock data predict slightly more positive signals 
than the observational data. 
A cause of this small discrepancy might be a simplicity of our source model. 
In our source model, the number density of UHECR sources is assumed 
to be constant in the whole Universe. 
If the number density of UHECR sources at a distant universe 
is relatively larger than at the local universe, 
cross-correlation between the distribution of local galaxies 
and the arrival directions of UHECRs is expected 
to become smaller, 
and may be fitted to that calculated from the observed data better. 
Although we would like to construct an improved source model 
in which a radial profile of the source number density is taken into account, 
we have little knowledge of UHECR sources. 
Thus, it will be studied when the number of detected events 
increases in the northern hemisphere 
and more information on UHECR sources can be obtained. 
The energy-scale might be a cause too 
because cosmic rays with lower energies can reach the earth 
from more distant sources. 
This gives a sense that correlation with nearby galaxies is weaken.

We also investigate cross-correlation 
between all-sky arrival distribution of 
both the PAO and AGASA data, and the IRAS galaxies with $z \leq 0.018$. 
For the combination of the two experimental data, 
the difference of energy-scale must be corrected. 
The energy spectra reconstructed by several experiments are different 
at the highest energy range (see Fig.5 in Ref. \cite{berezinsky08}) 
since UHECR experiments has a systematic error of $\sim 30\%$ 
on energy determination. 
Following Ref.\cite{berezinsky05}, a dip-calibration method, 
we shift UHECR energies of the AGASA data by $10\%$ to lower energies. 
The dip-calibration method requires the injection spectrum 
of protons with $\alpha = 2.6$, which is used throughout this section. 
On the other hand, the energy spectrum of the PAO is not consistent with 
the calibrated spectra of the other experiments 
even if the energies are shifted maximally 
within systematic energy errors ( 22\% \cite{dawson08}) \cite{berezinsky08}. 
Thus, we shift the PAO data by 50\% to higher energies 
to be consistent with the shifted flux 
of the other experiments \cite{berezinsky08}. 
Then, we adopt $8.55 \times 10^{19}$ eV 
as an energy threshold of events used, 
which corresponds to $5.7 \times 10^{19}$eV 
in the original energy-scale of the original PAO. 
As a result, we use all of the 27 PAO data 
and 9 events of the AGASA events 
whose original energies are above $9.5 \times 10^{19}$eV. 
The 9 events are shown in red in Fig. \ref{fig:map}. 
Note that in a recent paper by the PAO their energy scale is slightly corrected 
to higher energies slightly \cite{pao08c}, 
but the event data with the energies have not been published yet. 
Thus, we do not use the new energy scale.

\begin{figure}[t]
\begin{center}
\includegraphics[width=0.95\linewidth]{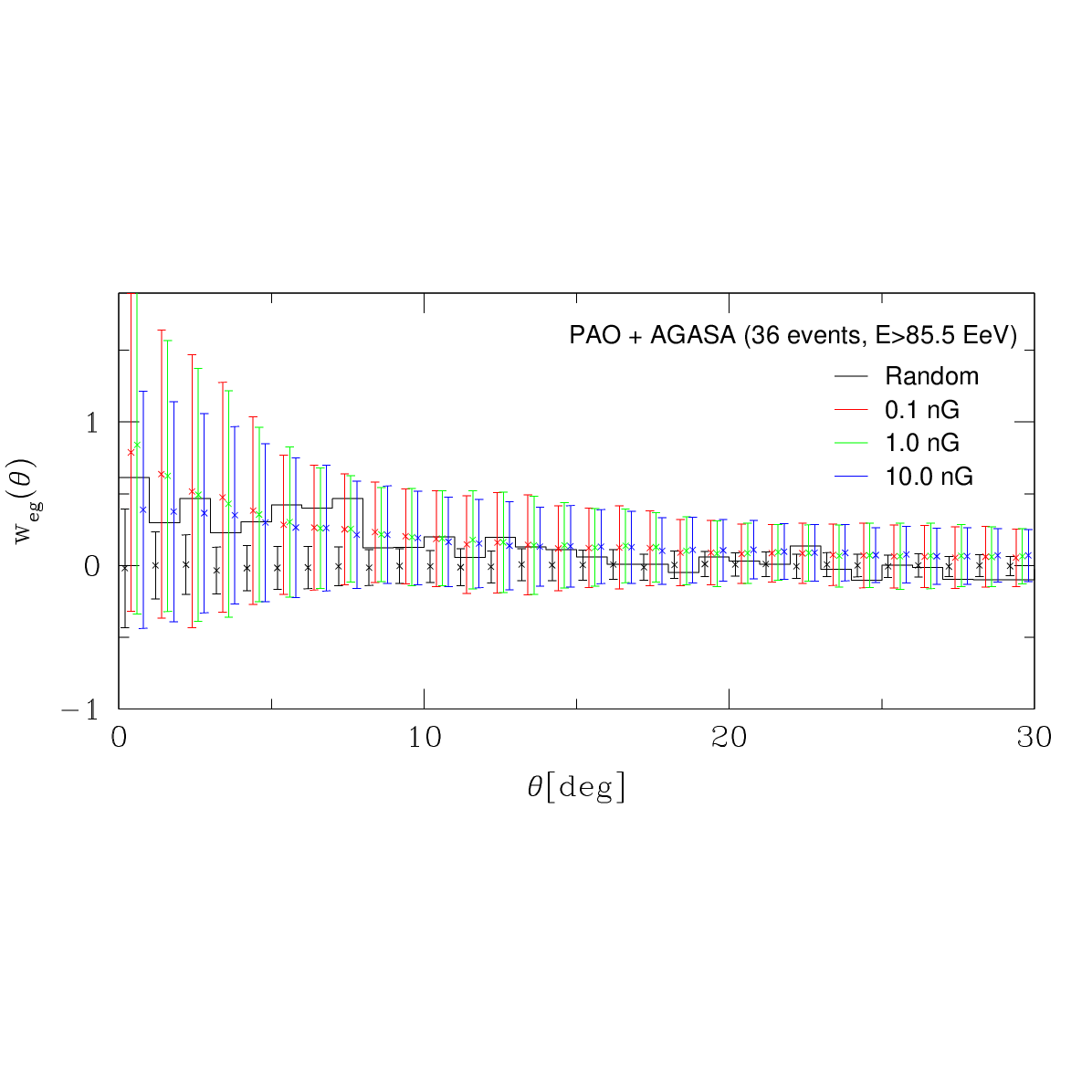}
\caption{The same as Fig.\ref{fig:model1}, but for the combined events.}
\label{fig:model2}
\end{center}
\end{figure}

Fig.\ref{fig:model2} represents cross-correlation functions 
calculated from the combined data ({\it histogram}), 
randomly distributed events ({\it black}), and mock UHE protons ({\it colors}). 
Comparing the histogram and black data 
(the same discussion in Figs.\ref{fig:ccor_pao} and \ref{fig:ccor_agasa}), 
we find a significant positive correlation signal within $8^{\circ}$. 
This angular scale is comparable with the angular scale of 
the auto-correlation of nearby galaxies in all sky, 
and thus the all-sky data also indicates 
that the distribution of UHECR sources is related to the distribution 
of local galaxies. 
The 3 IGMF model predictions are consistent with the observed data 
within the standard deviations. 
Although UHECRs in the northern hemisphere are still controversial, 
this all-sky analysis indicates that 
the observed highest energy cosmic rays are consistent 
with a scenario in which their source distribution 
is associated with nearby galaxy distribution.

Finally, we comment on a possibility that the large error bars are reduced 
and we can constrain the IGMF strength by future observations. 
The error bars originate from both the finite number of observed events 
and the sampling of galaxies. 
The former error can decreases by increasing the number of events, 
while the latter error does not decrease. 
We checked the possibility by simulating 200 mock protons above $\sim 60$ EeV. 
As a result, $B=10.0$ nG and $B=1.0$ nG are not distinguishable 
because of the latter error is left.

\section{Discussion \& Conclusion} \label{conclusion}

In this paper, 
we have investigated spatial correlation 
between the arrival directions of the highest energy cosmic rays 
observed by the PAO and AGASA, 
and the IRAS galaxies 
as a representative of the local structure of the universe. 
For statistical analysis, 
we adopted a cross-correlation function 
which can find the angular scale of the spatial correlation. 
This statistic need not to be set an artificial angular scale 
on correlation studies. 
We confirmed that 
the arrival directions of UHECRs above $5.7 \times 10^{19}$eV 
detected by the PAO are inconsistent with isotropic distribution 
and found that these have significant correlation with the galaxy distribution 
in the local universe inside $z=0.018$ 
within the angular scale of $\sim 15^{\circ}$. 
On the other hand, 
the AGASA data did not have obvious correlation with the IRAS galaxies. 
We also performed the same cross-correlation analysis on the whole sky 
using the combined data set of both the PAO and AGASA, 
and found the evidence of positive correlation 
within the angular scale of $\sim 8^{\circ}$. 
These results could be well reproduced by a source model 
that UHECR sources are distributed related to galaxies 
and the source number density is $\sim 10^{-4}$ Mpc$^{-3}$, 
taking propagation process in magnetized intergalactic space into account. 
These indicate that the distribution of UHECR sources are related 
to that of galaxies.

An important point of the results in this study is 
the angular scale in which the correlation is positive. 
We showed that the combined data has positive correlation 
compared to random event distributions within $\sim 8^{\circ}$ 
in Section \ref{modelp}. 
This scale is comparable with the angular scale of 
the auto-correlation of galaxies within $z=0.018$. 
Thus, we cannot distinguish the origin of the cross-correlation scale; 
the deflection angles of UHECRs in the IGMF 
or the spread of UHECR source distribution, 
but the result provides us with an upper limit of $\theta_{\rm obs}$. 
It enables us to constrain several structured IGMF models. 
Several simulations of the LSS formation with magnetic field 
have been performed 
and have applied to the studies of UHECR propagation 
in the local universe \cite{sigl04,dolag05,das08}, 
but the resultant magnetic structures are different 
except the centers of clusters. 
Among the three simulation results, 
70\% of protons reach the Earth with $\theta_{\rm obs} \geq 15^{\circ}$ 
even for the energies above $10^{20}$ eV 
in an IGMF model of Ref. \cite{sigl04}, 
in which only sources are strongly magnetized. 
Such large deflection angles are inconsistent 
with the angular scale of the spatial correlation 
estimated in this study. 
Thus, the magnetic structure in local Universe proposed by Ref. \cite{sigl04} 
is not consistent with the PAO data. 
Note that the authors of Ref. \cite{sigl04} pointed out the uncertainty 
of observer positions in their simulated universe 
(in other words, the magnetic structure of local Universe). 
The angular scale gives information on a magnetic field in local Universe

The angular scale of positive correlation against random event distribution 
also have information on the composition of UHECRs at the Earth, 
though interpretation depends on the results of the LSS formation simulations. 
Since UHECR sources are expected to be associated with dense structures, 
like clusters of galaxies and filamentary structures, 
propagating UHECRs are affected by magnetic fields in these structures. 
A LSS simulation by Ref. \cite{das08} showed 
that the filamentary structure has relatively 
strong magnetic field up to 10 nG. 
A simple but structured IGMF model developed in Ref. \cite{takami06} 
also reproduce such structures. 
Propagating protons above $6 \times 10^{19}$ eV 
in the magnetic structure by Ref. \cite{das08} 
is deflected by less than $\sim 10^{\circ}$, 
which is comparable with the angular correlation scale. 
When we assume heavy composition of UHECRs, like irons, 
the deflection angles of UHECRs are a few tens times larger than protons. 
This conflicts with our estimated angular correlation scale. 
Thus, protons or light nuclei are favored 
as the composition of the highest energy cosmic rays 
if the IGMF models proposed by Refs. \cite{das08,takami06} are real. 
On the other hand, a simulation by Ref. \cite{dolag05} predicted 
the filamentary structure with the IGMF strength of 0.1-1 nG 
and the volume fraction with the strength of magnetic field 
larger than 1nG is much smaller than the other 2 simulations. 
Thus, heavy nuclei-dominated composition is allowed 
if the IGMF model of Ref. \cite{dolag05} is real.

Another approach to estimate the strength of intervening magnetic field 
is the measurement of the separation angle 
between real UHECR sources and the arrival directions of UHECRs from them. 
The authors of Ref. \cite{angelis08} estimated 
the IGMF strength by using the correlation scale proposed by PAO 
($\sim 3.1^{\circ}$) as sub-nG 
assuming the PAO-correlated AGNs are real UHECR sources, 
though the assumption itself is controversial at present. 
When several sources are identified in the future, 
more detailed approaches to the magnetic field can be performed 
like Ref. \cite{angelis08}.

In fact, it has shown that 
the GMF significantly contributes to the deflection angles of UHECRs 
\cite{takami06,stanev97,medinatanco98,alvarez02,kachelriess07,yoshiguchi04,takami07b}. 
Ref. \cite{takami07b} showed that the trajectories of UHE protons with 
$\sim 6 \times 10^{19}$ eV are deflected by $\sim 4^{\circ}$ on average 
in BS models except for protons arriving from the direction of 
the Galactic center. 
The deflection pattern is complicated and dependent on the arrival directions 
of protons. 
Also, GMF could affect the arrival directions of protons 
not only arriving near the Galactic plane 
but also arriving far from the Galactic plane \cite{takami07b}. 
In this situation, we completely neglected GMF 
and adopted all data without removing events with arrival directions 
near the Galactic plane as a first step in this study. 
The statistical analysis taking GMF into account is a next target of our study.

Finally, we comment on the calibration of the energy-scale. 
When we discussed the cross-correlation using the combined data, 
the energies of the AGASA data were shifted by 10\% to lower energies 
and those of the PAO data were shifted by 50\% to higher energies 
on the assumption of the dip-calibration, 
a physically motivated calibration method \cite{berezinsky08,aloisio07}. 
However, the energy-shift of the PAO data is larger than 
its systematic uncertainty of 22\% \cite{pao08,dawson08} 
while the shift of the AGASA is within its systematic uncertainty 
\cite{hayashida00,takeda03}. 
If the spectral dip at around $10^{19}$eV is generated 
by pair-creation interaction with the CMB, 
either/both experiments possibly underestimate the uncertainty 
on the energy determination. 
On the other hand, if we assume that the energy-scale of the HiRes is correct, 
the energy spectra of all UHECR observatories can be consistent 
by shifting the energies of each experimental data 
within each systematic uncertainty \cite{berezinsky08b}. 
In this standpoint, 
we can interpret the spectral ankle not as the pair-creation dip 
but as the transition point from Galactic to extragalactic cosmic rays. 
The precise determination of the position of the dip enables us 
to distinguish the two interpretations on the ankle. 
When the correlation with extragalactic objects is investigated 
with combined data sets of observations 
of both the northern and southern hemispheres, as we have done, 
the calibration of the energy-scale is important 
because the GZK radius of protons in energy range 
from $6 \times 10^{19}$ to $10^{20}$ eV is sensitive to their energy. 
Precise energy determination is required 
in order to develop charged particle astronomy. 
A consensus between the energy-scales of the experiments is required.

Now the PAO and Telescope Array \cite{fukushima07} are operating 
and Extreme Universe Space Observatory (JEM-EUSO) \cite{ebisuzaki07} and 
the northern site of the PAO \cite{nitz07} are proposed. 
These experiments will not only accumulate much more statistics 
of UHECR events, but also determine the energy-scale precisely 
by more understanding physics in extensive air shower. 
We will be able to discuss UHECR sources more precisely in the near future.

\ack
We are grateful to Susumu Inoue and Tokonatsu Yamamoto for useful discussions. 
The works of H.T., T.N, and K.Y. are supported 
by Grants-in-Aid from JSPS Fellows. 
The work of K.S. is supported by Grants-in-Aid for 
Scientific Research provided 
by the Ministry of Education, Culture, Sports, Science and Technology (MEXT) 
of Japan through Research Grants S19104006. 
This work was also supported by World Premier International 
Research Center Initiative (WPI Initiative), MEXT, Japan.

\newpage

\end{document}